\documentclass[%
 preprint,
 amsmath,amssymb,
 aps,
pra,
]{revtex4-2}

\usepackage{graphicx}% Include figure files
\usepackage{dcolumn}% Align table columns on decimal point
\usepackage{bm}% bold math
\usepackage{array}

\begin{document}

\preprint{IUPAP CCP2024 Proceedings}

\title{Droplet shape classification in latent space}
\title{Representation of Typical Droplet Shapes in 2D Latent Space}

\author{Mihir Durve$^{1}$, Jean-Michel Tucny$^{1,2}$, Andrea Montessori$^{2}$, Marco Lauricella$^{3}$, and Sauro Succi$^{1,4}$}

\affiliation{%
$^1$Center for Life Nano- \& Neuro-Science, Italian Institute of Technology (IIT), viale Regina Elena 295, Rome, 00161, Italy\\
$^2$Dipartimento di Ingegneria, Università degli Studi Roma tre, via Vito Volterra 62, Rome, 00146, Italy\\
$^3$Istituto per le Applicazioni del Calcolo del Consiglio Nazionale delle Ricerche, via dei Taurini 19, Roma, 00185, Italy\\
$^4$Department of Physics, Harvard University, 17 Oxford St, Cambridge, MA 02138, United States
}

\begin{abstract}
In this study, we investigate the clustering of 5000 droplets, each originating from one of five distinct droplet classes, each representing a unique geometry. The shape coordinates of the droplets are mapped to a 2D latent space through a two-step, fully reversible process involving Fourier series and autoencoders. Thus, each droplet contour sampled by 400 points is represented by 2 scalar numbers.  We present data mapping 5000 droplets in two-dimensional space, which reveals ring-like clusters. Each cluster corresponds to a distinct droplet class, with the classification based on the droplet morphology, which is known a priori. Further, we find that well-known clustering algorithms like DBSCAN show limited success in identifying these ring-shaped clusters underscoring the need to explore alternative clustering approaches, particularly those tailored to handle ring-shaped distributed data. 
\end{abstract}

\keywords{Autoencoders, Microfluidics, Droplets, Latent space}

\maketitle

\section{Introduction}
Droplet generation is a cornerstone of numerous industrial processes, including the production of gels, emulsions, pharmaceuticals, and food products \cite{CHEN,Schroen,Chen2022}. In these applications, precise control over droplet properties, such as size and shape, are vital for ensuring product quality, uniformity, and functionality. For example, in emulsion and paint production, the consistency of droplet size and morphology significantly impacts key product characteristics, including viscosity, texture, stability, and even aesthetic attributes. Consequently, the ability to generate droplets with tightly controlled physical and chemical properties is a critical requirement in these industries.

The task of producing droplets with specific and precise characteristics poses a complex challenge \cite{Ding,Lathia}. Droplet formation is influenced by a variety of factors, such as the composition of liquid mixtures, temperature, surface tension, flow dynamics, and droplet velocity. Each parameter must be meticulously tuned to achieve the desired droplet properties. Despite advances in droplet generation technologies, achieving precise and reproducible control over these variables remains a challenging task. In industrial settings, this optimization process is often performed manually, relying on manual adjustments informed by visual inspection and discrepancies between observed and target droplet attributes. Such manual approaches are cost-ineffective, time-consuming, and prone to error, particularly in large-scale productions.

To overcome these limitations, there is increasing interest in the deployment of automated, self-learning algorithms for optimizing droplet generation. Machine learning techniques, especially reinforcement learning, have emerged as promising tools for replacing manual interventions \cite{Dressler,Gyimah}. These algorithms can autonomously adjust the process parameters by learning from the feedback of the system, thus minimizing human involvement and improving operational efficiency. However, for machine learning models to effectively control droplet production, it is essential to represent the complex geometries of droplets in a low-dimensional space.

In recent works \cite{durve2024first,durve2024}, we introduced a two-step methodology to map droplet shapes to a two-dimensional latent space, facilitating the application of reinforcement learning to droplet production systems. In the first step, the geometry of each droplet is described using a Fourier series expansion with a finite number of terms. This approach captures the essential features of droplet shapes in a compact functional form, leveraging the Fourier series’ ability to efficiently represent closed loop geometries with relatively few coefficients. In the second step, these Fourier coefficients are transformed into a two-dimensional latent space using an autoencoder neural network. Autoencoders are particularly well-suited for dimensionality reduction tasks, as they can learn compact, informative representations of high-dimensional data. This approach enables seamless integration of machine learning models with droplet generation systems, paving the way for automated, high-precision control in industrial processes.

In this paper, we present an analysis of typical droplet shapes as represented in a two-dimensional (2D) latent space. Our findings reveal that these droplet shapes organize into distinct, ring like clusters within the latent space, providing a visually intuitive and structured representation of their geometric variations. The paper is structured as follows: In the next section \ref{method}, we provide a summary of the two-step procedure used to map droplet shapes into the 2D latent space. Then, in sec. \ref{lp}, we present the results of our latent space mapping for nine distinct types of droplets obtained from experimental observations. Finally, in the last sec. \ref{Conclusion}, we conclude the paper by summarizing our key findings and discussing the broader implications of these results.

\section{Method to compute droplet shape coordinate in latent space}
\label{method}

The method used to map a given droplet contour onto a 2D space involves a two-step procedure. In the first step, the droplet contour is represented using a finite-term Fourier series in complex notation. Typically, 21 Fourier modes are sufficient to capture the intricate and extreme droplet shapes observed in high-density emulsions. In the second step, an autoencoder neural network is employed to map these Fourier series coefficients to just two real numbers, allowing the model to learn a low-dimensional representation (latent space) of the droplet shapes. This process is fully reversible; that is, the droplet contour can be recovered from the latent space representation. In the reverse process, the autoencoder predicts the Fourier series coefficients from the latent space coordinates, and the droplet contour is then reconstructed by simply computing the droplet contour coordinates from the Fourier series using the predicted coefficients. A detailed description of the procedure can be found in Ref.\cite{durve2024}, but for completeness, we provide a summary here.

\subsection{Step I - Fourier series representation of droplet contour}

The individual droplet contour is represented by approximately 400 x,y coordinate pairs. These droplets are selected from an experimental setup designed to investigate droplets in dense oil-water emulsions (see Ref. \cite{durve2024}). The contour traced by these x,y coordinates is denoted as $f_o(t)$, where the parameter $t$ is within the interval [0,1]. The Fourier series that approximates $f_o(t)$ over this interval, with a finite number of terms, is expressed as:

\begin{equation}
 f_o(t) = \sum_{n=-\infty}^{+\infty} c_{n} e^{n \cdot 2\pi i t},
 \label{eq1}
\end{equation}

where $n$ represents the number of modes in the Fourier series, and $c_n$ are the complex Fourier coefficients. It is found that $n=10$ is sufficient to accurately represent the extreme droplet shapes observed in the experiments.

The Fourier coefficients $c_n$ are computed as follows:

\begin{equation}
 c_n = \int_{0}^{1} f_o(t) e^{-n \cdot 2 \pi i t} dt \approx \sum_{t=0}^{1} [f_o(t) e^{-n \cdot 2 \pi i t} \Delta t],
 \label{eq2}
\end{equation}

where the time step $\Delta t$ is defined as $\Delta t =$ 1.0/(number of points on the contour). Equations \ref{eq1} and \ref{eq2} provide a straightforward method for computing the Fourier series coefficients from the droplet contour and vice versa. For $n=10$, we obtain 21 complex coefficients, each of which corresponds to 42 real numbers in the form $x + iy$, where $x$ and $y$ represent the real and imaginary parts, respectively.

In the second step, these 42 real numbers are mapped to 2 scalar numbers with the help of autoencoder networks as described below. 

\subsection{Step II - Autoencoders}

We set up an autoencoder network that takes the 42 Fourier coefficients as input and produces an identical output. The network hyperparameters are detailed in Table \ref{table}. The autoencoder is trained to replicate the input as the output. However, the data passes through the bottleneck layer, which consists of only 2 neurons. After training, the values of these 2 neurons at the bottleneck layer contain sufficient information to predict the output. This dimensionality reduction process maps the 42 real numbers to just two numbers, effectively reducing the approximately 400 data points representing the droplet contour to 42 real numbers corresponding to the Fourier series coefficients, and then further mapping those coefficients to 2 real numbers.

\begin{table}[h!]
\centering
\begin{tabular}{ |p{5cm}||p{6cm}|}
 \hline
 Activation function  & PReLU    \\
 \hline
 Optimizer  & Nadam    \\
 \hline
 Learning rate&  \parbox{5cm}{Epoch $< 15$ : $2 \times 10^-3$\\ Epoch 15-20 : $2 \times 10^-4$\\ Epoch 21-25 : $1 \times 10^-5$ \\ Epoch $>$ 25: $1 \times 10^-6$ \\ }   \\
\hline
 Network configuration & \parbox{5cm}{ Input: 42\\ Layer 1: 50\\ Layer 2: 100\\ Layer 3: 250\\ Layer 4: 250\\ Layer 5: 100\\ Layer 6: 10\\ Layer 7: 2 \\ Layer 8: 2\\ Layer 9: 2 \\ symmetric \\ .\\ .\\ .\\Output: 42 \\} \\
 \hline
 Number of trainable parameters & \parbox{5cm}{ 253687 } \\
  \hline
\end{tabular}
\caption{ \label{table} The autoencoder network configuration and hyperparameters (See Ref. \cite{durve2024} for operational details).}
\end{table}

Next, we investigate the latent space coordinates obtained through this method for droplets observed in dense emulsions.

\section{Clustering in latent space}
\label{lp}

\begin{figure}
    \centering
    \includegraphics[width=0.85\linewidth]{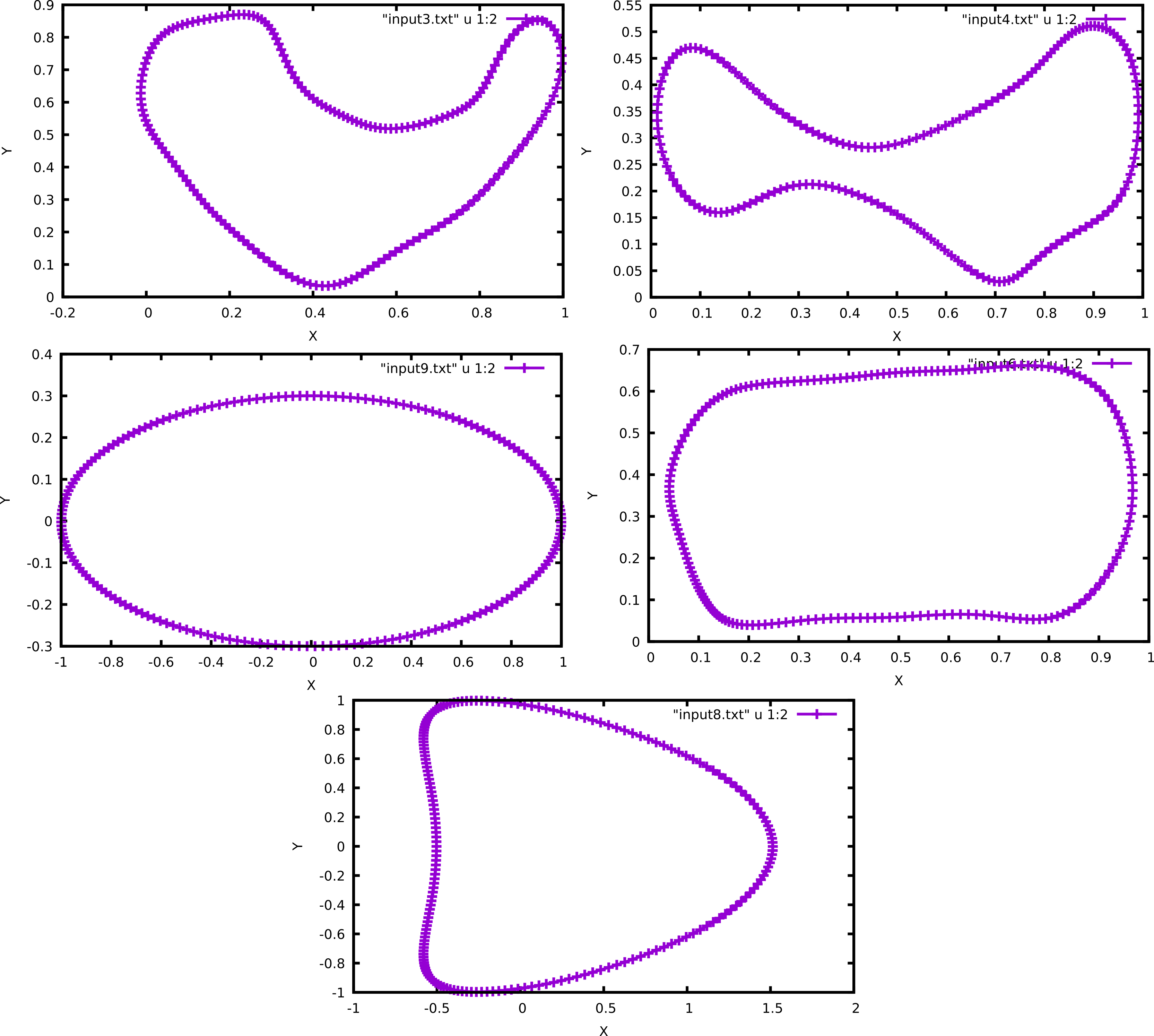}
    \caption{Representative droplet shapes from five distinct classes are presented. The classification of droplets was performed through visual inspection of experimental results obtained from oil-water emulsion generation using various microchannel geometries. \label{drop_class}}
\end{figure}

\begin{figure}
    \centering
    \includegraphics[width=0.75\linewidth]{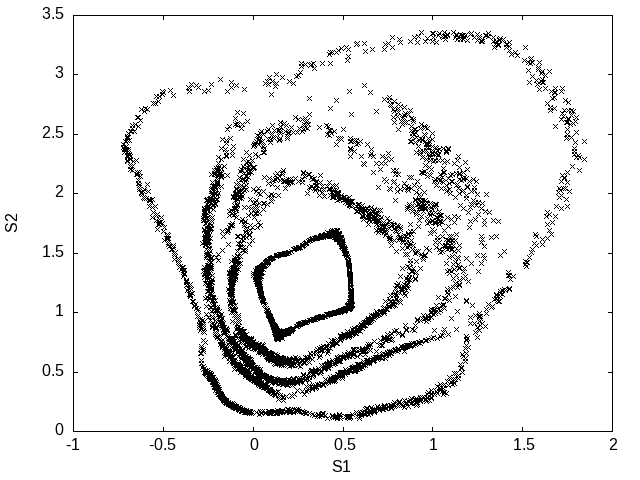}
    \caption{ Latent space representation of 5000 droplets taken from five droplet classes shown in Fig. \ref{drop_class}.
    \label{All_droplets}}
\end{figure}

\begin{figure}
    \centering
    \includegraphics[width=1\linewidth]{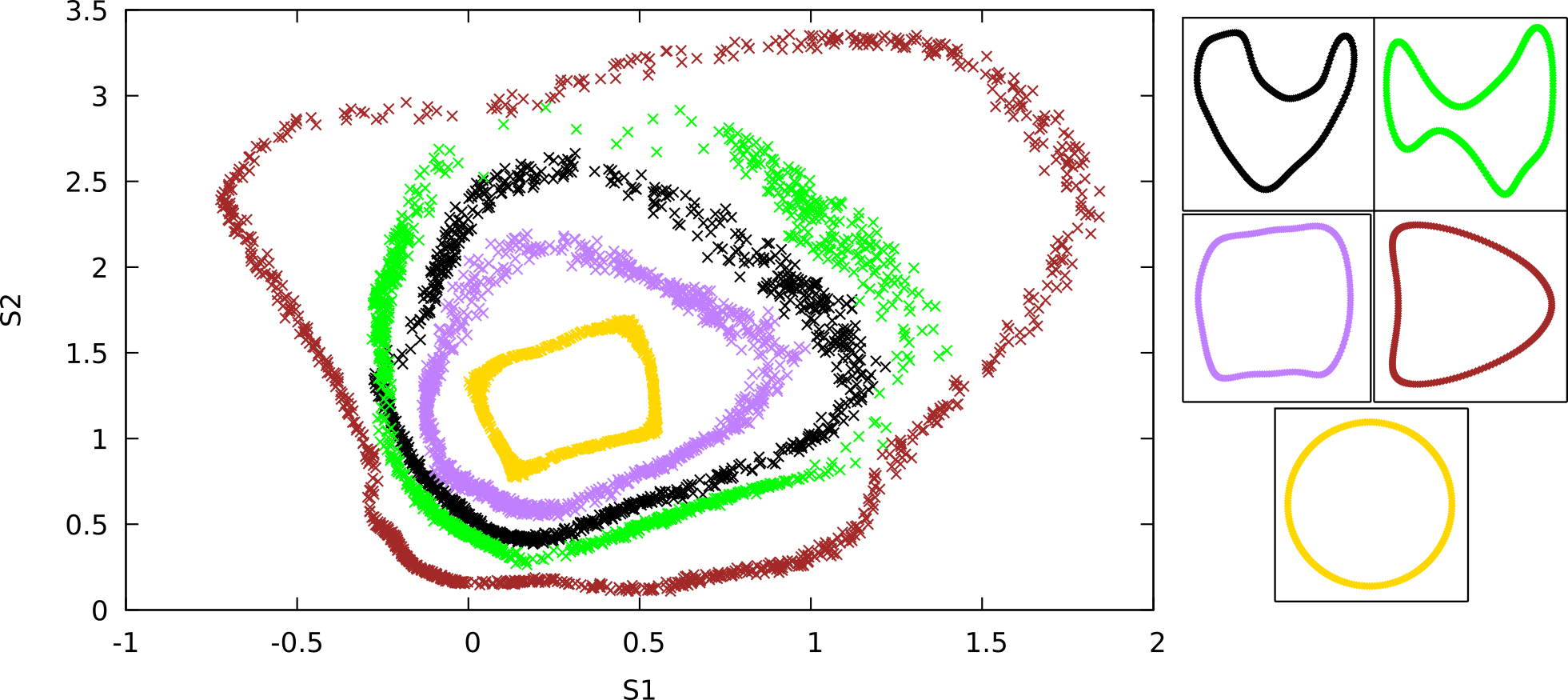}
    \caption{Latent space representation of 5000 droplets belonging to five distinct classes. The right-hand panel displays the representative shapes for each droplet class, distinguished by unique colors. Each point in the latent space corresponds to a mapped representation of a distorted variant of the droplet depicted in the right-hand panel.  
    \label{All_droplets_classes}}
\end{figure}

\begin{figure}
    \centering
    \includegraphics[width=1\linewidth]{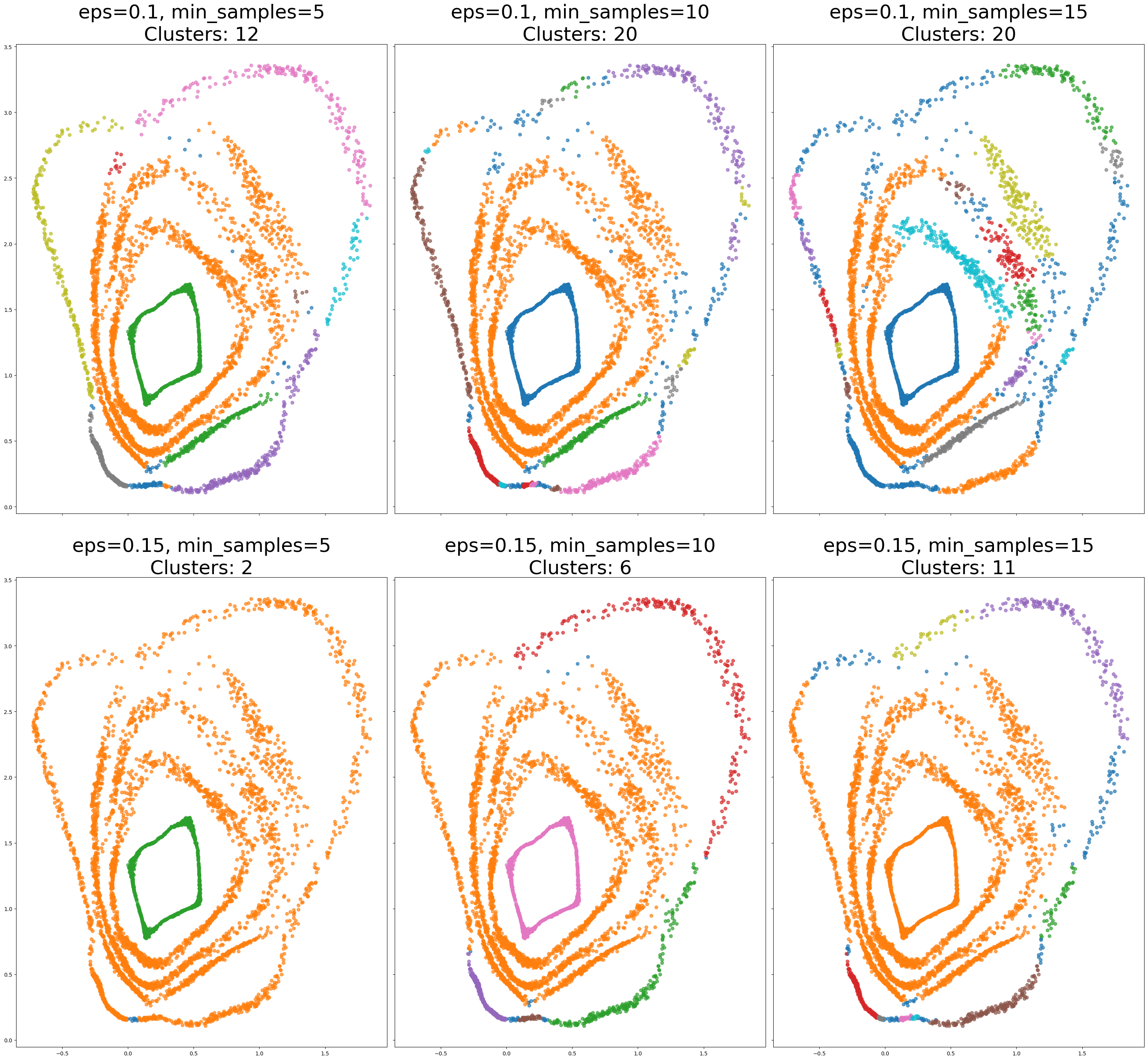}
    \caption{Results of DBSCAN Clustering algorithm for various values of the parameter eps and min\_sample. Each color represents a unique cluster.  
    \label{dbscan}}
\end{figure}

To investigate the mapping of droplet shapes into the 2D latent space, we utilize a dataset consisting of approximately 5000 droplets derived from experimental observations. The droplet shapes used in this study are already shown in our previous work (see Fig. 2 of Ref. \cite{durve2024}). This dataset includes around 1000 droplets from each of five distinct droplet classes (see Fig. \ref{drop_class}). The 1000 droplets within each class are generated by perturbing the basic droplet shape of the class. This is accomplished by introducing noise to all Fourier series coefficients that define the basic shape. The amplitude of the noise is set to be no more than 0.1 times the value of each coefficient. These droplets are mapped to the 2D latent space using the two-step process described in Section \ref{method}. Upon completion of the mapping process, each droplet is represented by two real-valued parameters, denoted as S1 and S2.

In Fig. \ref{All_droplets}, we present the mapping of all 5000 droplets in the 2D latent space, where the axes correspond to the S1 and S2 parameters. The plot reveals five concentric rings, with occasional intersections between them. However, the visualization does not provide a clear indication as to whether each droplet class forms a distinct and separate ring within the latent space, as some overlap between the classes is evident.

In Fig. \ref{All_droplets_classes}, we present a refined version of the latent space mapping, where each droplet class is assigned a unique color for better differentiation. On the right-hand side of this figure, we include representative droplets from each class, with their corresponding colors, to observe the relationship between the latent space clusters and the original droplet geometries. Although the droplet classes exhibit somewhat distinct ring-like formations, there is noticeable overlap between the rings, which obscures the boundaries between the clusters of different droplet classes.

We employed DBSCAN (Density-Based Spatial Clustering of Applications with Noise) to identify the number of clusters in the data. DBSCAN (Density-Based Spatial Clustering of Applications with Noise) is a widely adopted density-based clustering algorithm that partitions data points according to their spatial proximity and local density \cite{ester1996}. Unlike centroid-based algorithms such as K-means, DBSCAN does not require the user to predefine the number of clusters. It classifies data points into three categories: core points (which have a minimum number of neighboring points, defined by the parameter "min\_samples", within a specified radius, "eps"), border points (which have fewer neighbors but lie within the neighborhood of a core point), and noise points (which do not belong to any cluster). Figure \ref{dbscan} illustrates the results of DBSCAN clustering for various parameter settings. Despite its advantages, the algorithm shows limited success in accurately identifying the correct number of clusters when applied to ring-shaped data, as seen in Fig. \ref{All_droplets}.

We performed a three-stage analysis to characterise droplet shapes, and we have evaluated the computational time required for this process. Initially, the YOLOv11 segmentation model is utilised to accurately delineate droplet boundaries from input images. Then the acquired droplet shapes are used to generate additional droplet shapes in the same class using a Fourier series with added noise. In the second stage, an autoencoder maps the extracted shape coordinates onto a two-dimensional (2D) latent space. Finally, in the third stage, a clustering algorithm groups the 2D coordinates into distinct clusters.

The computational time for each stage was measured using a mid-range laptop equipped with an Intel i7 processor (6 Performance-cores and 14 Efficient-cores) and 32 GB of RAM. The results are averaged over 5000 instances and are summarised in the table below. It is worth noting that the CPU computational times mentioned below are indicative of the real-world test at best, and they are subject to the computing hardware as well as the software used for the analysis. 

\begin{table}[h!]
\centering
\begin{tabular}{|p{6cm}|p{6cm}|}
\hline
\textbf{Stage} & \textbf{ Computational Time (ms)} \\
\hline
\parbox{6cm}{1: YOLOv11 segmentation} & \hspace{2cm} 1458 \\
\hspace{3cm} or & \\
\parbox{6cm}{1: Droplet shape generation} & \hspace{2cm} 11 \\ 
\hline
\parbox{6cm}{2: Autoencoder Mapping} & \hspace{2cm} 53 \\
\hline
\parbox{6cm}{3: Clustering} & \hspace{2cm} 23 \\
\hline
\end{tabular}
\caption{Computational time for the three-stage analysis for each droplet shape. The computing time is averaged for processing 5000 such shapes.}
\label{tab:comp_time_formatted}
\end{table}

\section{Conclusion}
\label{Conclusion}
In this work, we map about 5000 droplets from 5 distict classes of droplets on to a 2D latent space. The mapping was done by a two-step process, which involves representing the droplet boundary coordinates by finite term Fourier series as the first step and then mapping the limited Fourier series coefficients on a 2D space using autoencoder neural networks. The points in the 2D space formed cocentric rings with each ring corresponding to a unique class of droplets. These rings were not completely distict and showed occasional overlapping of a few data points. DBSCAN clustering algorithm showed limited success in identifying the correct number of clusters in this data. A suitable clustering algorithm such as Fuzzy clustering algorithm, especially designed for elliptically spaced clusters, by Gath and Hoorey \cite{gath1995} should be tested to identify the ring like clusters which we have identified manually in this work. If successful, the overall process can identify droplet morphology classes with unsupervised learning.

\section{Acknowledgment}
The authors acknowledge funding from the European Research Council ERC-PoC2 grant No. 101081171
(DropTrack) and grant No. 101187935 (LBFAST). J.-M. T. thanks the FRQNT “Fonds de recherche du Quebec – Nature et technologies (FRQNT)”
for financial support (Research Scholarship No. 314328). M.L. acknowledges the support of the Italian National
Group for Mathematical Physics (GNFM-INdAM). M.D. thank Adriano Tiribocchi for useful discussions.
\bibliography{version1}

\end{document}